\documentclass[superscriptaddress,prc,twocolumn,showpacs,preprintnumbers,altaffilletter,aps,showpacs,10pt,groupedaddress]{revtex4-1}
\usepackage{xcolor}       % use if color is used in text
\usepackage{graphicx}
\usepackage{setspace}
\usepackage{subfigure}
\usepackage{multirow}
\usepackage{amssymb,amsmath}
\usepackage{siunitx} %MF. Includes proper formating for SI units
\usepackage{datetime} %MF. To include the current time in the date stamp.
\usepackage{xspace}
\usepackage{hyphenat}
\usepackage{bbold}

\begin{document}
\title{Violation of the Leggett-Garg Inequality in Neutrino Oscillations}

\newcommand{\mitt}{Massachusetts Institute of Technology, Cambridge, MA, USA}

\newcommand{\comment}[1]{}

\author{J.\,A.~Formaggio}
\author{D.\,I.~Kaiser}
\author{M.\,M.~Murskyj}
\author{T.\,E.~Weiss}
\affiliation{\mitt}

\begin{abstract}
The Leggett-Garg inequality, an analogue of Bell's inequality involving correlations of measurements on a system at different times, stands as one of the hallmark tests of quantum mechanics against classical predictions. The phenomenon of neutrino oscillations should adhere to quantum-mechanical predictions and provide an observable violation of the Leggett-Garg inequality. We demonstrate how oscillation phenomena can be used to test for violations of the classical bound by performing measurements on an ensemble of neutrinos at distinct energies, as opposed to a single neutrino at distinct times. A study of the MINOS experiment's data shows a greater than $6\sigma$ violation over a distance of 735 km, representing the longest distance over which either the Leggett-Garg inequality or Bell's inequality has been tested. 
%By exploiting stationarity and the prepared-ensemble condition, rather than weak measurements, our results provide strong evidence against ``hidden-variable theories," which are deterministic alterives to quantum mechanics. 
\end{abstract}

\maketitle
\section*{Introduction}

Perhaps one of the most counterintuitive aspects of quantum mechanics is the principle of superposition, which stipulates that an entity can exist simultaneously in multiple different states. Bell and others indicated how experiments could distinguish between classical systems and those that demonstrate quantum superposition~\cite{Bell,Kochen}.
%This principle is not merely a speculative feature of quantum mechanics, but one that can be physically tested. Indeed, the propositions put forward by Bell~\cite{Bell} allow an experimenter to distinguish between classical systems and those that demonstrate quantum superposition~\cite{Kochen}.  
Bell's inequality concerns correlations among measurements on spatially separated systems. Leggett and Garg developed an analogous test that concerns correlations among measurements performed on a system at different times, and they extended this test to apply to macroscopic entities~\cite{LGI}. 
%Leggett and Garg derived an inequality that is compatible with characteristics of classical physics but is clearly violated in systems with quantum coherence~\cite{LGI}.  
Sometimes referred to as the ``time-analogue" of Bell's inequality, the Leggett-Garg inequality (LGI) allows for a complementary test of quantum mechanics while potentially avoiding some of the difficulties involved in performing a truly loophole-free test of Bell's inequality \cite{Hensen,NIST,Giustina,GFK}. See \cite{ELN} for a recent review.

The original goal of LGI tests was to demonstrate macroscopic coherence---that is, that quantum mechanics applies on macroscopic scales up to the level at which many-particle systems exhibit decoherence \cite{LGI,ELN,Waldherr,Zhou2012,Xu,Dressel}. For this reason, a major focus of recent LGI research has been scaling up to tests with macroscopic systems. 
%Notably, Zhou et al.~\cite{Zhou2012} recently reported finding LGI violation  caused by quantum coherence  in macroscopic crystals.

LGI tests have another purpose:~to test ``realism,'' the notion that physical systems possess complete sets of definite values for various parameters prior to, and independent of, measurement. ``Realism" is often encoded in hidden-variable theories, which allow for systems that are treated as identical according to quantum mechanics to be fundamentally distinguishable through a hidden set of parameters that they possess, such that any measurement on a system reveals a pre-existing value \cite{Huelga96}. LGI violations imply that such hidden-variable (or ``realistic'') alternatives to quantum mechanics cannot adequately describe a system's time evolution. Experiments using few-particle systems can test ``realism" even if they do not directly address \emph{macro}realism \cite{Hardy,Huelga95,Huelga96,Goggin, PalaciosLaloy,Athalye,Suzuki}.

Neutrino flavor oscillations, which are coherent in the few-particle limit, provide an interesting system with which to test the LGI. Neutrinos have been detected in three distinct ``flavors,'' which interact in particular ways with electrons, muons, and tau leptons, respectively. Flavor oscillations occur because the flavor states are distinct from the neutrino mass states; in particular, a given flavor state may be represented as a coherent superposition of the different mass states \cite{Camilleri,Duan}. Neutrino flavor oscillations may be treated with the same formalism that is typically used to describe other systems displaying quantum coherence, such as squeezed atomic states~\cite{Leroux:2010zz}. The major difference between neutrinos and these familiar systems, however, is that the coherence length of neutrino oscillations---the length over which interference occurs and oscillations may be observed---extends over vast distances, even astrophysical scales \cite{Jones15}. An LGI experiment using neutrino oscillations therefore presents a stark contrast to other types of LGI tests, which typically use photons, electrons, or nuclear spins, for which coherence distances are much more constrained \cite{ELN}.  

Experimental violations of the LGI can lead to definitive conclusions about ``realism'' only if the measurement outcomes represent the underlying time evolution of the system. Invasive measurements, characterized either by wavefunction collapse or by experimental imperfections that classically disrupt the system, would prevent an experimenter from ruling out realistic alternatives to quantum mechanics, even in the face of an apparent violation of the LGI. Several experiments have worked to bypass this limitation by using indirect or weak measurements to probe the system~\cite{Dressel,Goggin,PalaciosLaloy}. 

In the case of neutrino flavor oscillations, it is possible to circumvent the problems posed by invasivity by performing measurements on members of an identically prepared ensemble; this obviates the issue of whether individual measurements influence one another. When combined with a separate assumption of ``stationarity''---such that the correlations between different measurements depend only on the durations between them, rather than on their individual times---the prepared-ensemble condition allows one to test particular classes of ``realistic" alternatives to quantum mechanics \cite{ELN,Huelga95,Huelga96,Zhou2012}. Although the idea of testing the LGI and related measures of quantum entanglement using neutrino oscillations has been proposed in the literature \cite{Gangopadhyay,Alok2014,Banerjee2015}, we believe this is the first such empirical test to be performed.

\section*{Formalism and Assumptions}

We consider a dichotomic observable $\hat{Q}$ (with realizations $\pm 1$) that may be measured at various times $t_i$. The correlation between measurements at times $t_i$ and $t_j$ can be written
\begin{equation}
{\cal C}_{ij} \equiv \langle \hat{Q} (t_i) \hat{Q} (t_j) \rangle,
\end{equation}
where $\langle ... \rangle$ indicates averaging over many trials. For measurements at $n$ distinct times, we may define the Leggett-Garg parameter $K_n$ as
\begin{equation}
K_n \equiv \sum_{i = 1}^{n - 1} {\cal C}_{i, i+1}  -  {\cal C}_{n, 1} .
\end{equation}
``Realistic'' systems obey the Leggett-Garg inequality \cite{LGI,ELN}, which for $n\geq 3$ is given by $K_n \leq n-2$.
%\begin{equation}
%\begin{split}
%-n \leq &K_n \leq n - 2 \>\> {\rm for} \> n \> {\rm \> odd} , \\
%-(n - 2) \leq &K_n \leq n-2 \>\> {\rm for} \> n \> {\rm \> even}.
%K_n \leq n - 2 .
%\label{eq:LGI}
%\end{split}
%\end{equation}

We may calculate an expected value for $K_n$ according to quantum mechanics, $K_n^Q$, by time-evolving $\hat{Q}$ under the unitary operator ${\cal U} (t)$: $\hat{Q} (t_i) = {\cal U}^\dagger (t_i) \> \hat{Q} \> {\cal U} (t_i)$. When it is possible to represent $\hat{Q} (t_i)$ as $\hat{Q} (t_i) = \vec{b}_i \cdot \vec{\sigma}$, where $\vec{\sigma} = (\sigma_x , \sigma_y, \sigma_z)$ is a vector of two-by-two Pauli matrices, then for any complete set of normalized states 
%a vector sum of two-by-two Pauli matrices, then we may always write $\hat{Q} (t_i) = \vec{b}_i \cdot \vec{\sigma}$ and $\hat{Q} (t_j) = \vec{b}_j \cdot \vec{\sigma}$ (for unit vectors $\vec{b}_i , \vec{b}_j$, and where $\vec \sigma = (\sigma_x, \sigma_y, \sigma_z)^T$). For any complete set of normalized states 
$\vert \phi \rangle$, the two-time correlation function may be written
\begin{equation}
{\cal C}_{ij} = \frac{1}{2} \langle \phi \vert \{ \hat{Q}(t_i) , \hat{Q} (t_j) \} \vert \phi \rangle = \vec{b}_i \cdot \vec{b}_j.
\label{eq:cij}
\end{equation}
We use the anticommutator $\{ \hat{A}, \hat{B} \} \equiv \hat{A} \hat{B} + \hat{B} \hat{A}$ to avoid possible time-ordering ambiguities, since $\hat{Q}(t_i)$ need not commute with $\hat{Q} (t_j)$. 
%Then $K_n^Q$ becomes \cite{ELN}
%\begin{equation}
%K_n^{Q} =  \sum_{i=1}^{n-1} \cos (\vartheta_i) - \cos \left( \sum_{i = 1}^{n - 1} \vartheta_i \right),
%\label{eq:quantumK}
%\end{equation}
%where $\vartheta_i \equiv {\rm arccos} (\vec{b}_i \cdot \vec{b}_{i + 1} )$.

In a quantum-mechanical system, violations of the LGI arise due to the nonvanishing commutators of the operators $\hat{Q} (t_i)$ and $\hat{Q}(t_j)$. If one artificially imposes that $\hat{Q}(t_i)$ and $\hat{Q}(t_j)$ must commute (i.e., taking the limit $\hbar \rightarrow 0$), then one recovers the classical prediction for $K_n$, which we denote $K_n^C$, and which has the compact expression
\begin{equation} 
K_n^{C} = \sum_{i=1}^{n-1}  {\cal C}_{i,i+1} - \prod_{i=1}^{n-1}  {\cal C}_{i,i+1} .
\label{eq:classicalK}
\end{equation}
Given that ${\cal C}_{ij}$ is real and $\vert {\cal C}_{ij} \vert \leq 1$, we see that $K_n^C$ satisfies the LGI, whereas $K_n^Q \leq n \cos (\pi / n)$ may violate the LGI for particular angles $\vartheta_i \equiv {\rm arccos} (\vec{b}_i \cdot \vec{b}_{i + 1})$. The discrepancy between these two predictions provides an opening for experimental testing and verification~\cite{ELN}.

The original derivation of the LGI assumed that measurements of $\hat{Q}$ at various times $t_i$ are made in a non-invasive manner \cite{LGI}. The LGI may be derived instead under the assumption of ``stationarity,'' such that the correlation functions ${\cal C}_{ij}$ depend only on the time difference $\tau \equiv t_j - t_i$ between measurements \cite{Huelga95,Huelga96,Zhou2012,ELN}. In this case, the bound $K_n \leq n - 2$ applies to the class of ``realistic'' models that are Markovian, for which the evolution of the system after some time $t$ is independent of the means by which the system arrived in a given state at $t$ \cite{ELN}. We may then consider measurements performed on distinct members of an identically prepared ensemble, each of which begins in some known initial state. 

%We require two conditions in order to fulfill this assumption. First, we consider measurements performed on distinct members of an identically prepared ensemble, each of which begins in some known initial state. Second, we assume ``stationarity,'' that is, that the correlation functions ${\cal C}_{ij}$ depend only on the time difference $\tau \equiv t_j - t_i$ between measurements.

Stationarity allows for measurements made on distinct ensemble members to mimic a series of measurements made on a single time-evolving system. For example, in order to construct $K_3$, we take advantage of the fact that ${\cal C}_{23}$ in one system is equivalent to the correlation between measurements separated by time $\tau = t_3-t_2$ on a different member of the ensemble. 
%As a result, if the prepared-ensemble and stationarity conditions are satisfied, it is not necessary to make more than one measurement on a given system in order to test the LGI. 

The combination of the prepared-ensemble and stationarity conditions therefore acts as a substitute for measurement schemes intended to be non-invasive (e.g., weak measurements), because wavefunction collapse and classical disturbance in a given system do not influence previous or subsequent measurements on distinct members of the ensemble~\cite{Waldherr,Huelga96}. Unlike the assumption of non-invasive measurability, moreover, the stationarity condition may be subjected to independent testing \cite{Huelga95,Huelga96,Zhou2012}.  As we will see, both the prepared-ensemble and stationarity conditions may be fulfilled in measurements of neutrino flavor oscillations.  Furthermore, these two conditions enable us to analyze measurements on separate groups of particles (directly analogous to measurements on spatially separated systems in tests of Bell's inequality), circumventing the recent criticism of the LGI whereby measurements on a single system at later times may be influenced by the outcomes of earlier measurements on that same system~\cite{ClementeKofler}.

\vspace{-0.2cm}
\section*{LGI Violation Using Neutrinos}

The Standard Model includes three distinct neutrino flavors. However, the energies and distances on which we focus single out oscillations almost entirely between two flavor states. Hence, we adopt a two-state approximation. In the relativistic limit, oscillations between these two states may be treated with the Bl\"{o}ch sphere formalism \cite{Mehta09}, which geometrically represents the space of pure states of a generic two-level system. The observable $\hat{Q}$ measures neutrino flavor as projected along a particular axis (which we take to be $\hat{z}$): $\hat{Q} \equiv \sigma_z$, with eigenvalues $\hat{Q} \vert \nu_\mu \rangle = \vert \nu_\mu \rangle$ and $\hat{Q} \vert \nu_e \rangle = - \vert \nu_e \rangle$, for muon- and electron-flavor neutrino states, respectively.%\footnote{For the experimental data we consider below, $\vert \nu_e \rangle$ really represents not-$\vert \nu_\mu \rangle$, and might include some tiny admixture of tau neutrinos.}

The Hamiltonian for neutrino propagation in the two-flavor limit is given by (setting $\hbar = c = 1$) \cite{Camilleri,Duan}
\begin{equation}
\begin{split}
{\cal H} &= \left( p + \frac{ m_1^2 + m_2^2}{4p} + \frac{V_C}{2} + V_N \right) \mathbb{1}\\
&\quad\quad + \frac{1}{2} \left( \begin{array}{cc} V_C - \omega \cos 2\theta & \omega \sin 2 \theta \\ \omega \sin 2 \theta & \omega \cos 2 \theta - V_C \end{array} \right) \\
&\equiv r_0 \mathbb{1} + \frac{ \vec{r} \cdot \vec{\sigma}}{2} ,
\end{split}
\label{eq:H}
\end{equation}

\noindent where $\theta$ is the neutrino  vacuum mixing angle, $m_1$ and $m_2$ label the distinct mass states, $\omega \equiv (m^2_2 - m^2_1)/2p$ is  the oscillation frequency, and $p \simeq E$ is the relativistic neutrino momentum/energy.  The term $V_{C(N)} = \sqrt{2}G_F n_{e(n)}$ is the charged (neutral) current potential due to coherent forward scattering of neutrinos with electrons (neutrons) in matter, and $G_F$ is the Fermi coupling constant.  The term in Eq.~(\ref{eq:H}) proportional to $\mathbb{1}$ affects all flavor states identically and therefore does not contribute to flavor oscillations. %When treating evolution through matter, it is possible to define a new mixing angle $\theta_M$ such that $\hat{r} \equiv \vec{r}/|\vec{r}| =  (\sin 2\theta_M, 0 , -\cos 2\theta_M)$. Note that the vacuum oscillation Hamiltonian is recovered in the limit $V_C \rightarrow 0$.

For neutrinos of a given energy $E_a$, the time evolution of flavor states is governed by the unitary operator ${\cal U}$, which is related to ${\cal H}_{\rm osc} \equiv \vec{r} \cdot \vec{\sigma} / 2$ via
\begin{equation}
\begin{split}
{\cal U} (\omega_a; t_i, t_j ) &\equiv {\cal U} (\psi_{a; ij}) =  \exp \left[ - i \int_{t_i}^{t_j} {\cal H}_{\rm osc} (\omega_a) dt \right]   \\
& \simeq  \cos (\psi_{a; ij} )\mathbb{1} - i \sin (\psi_{a;ij }) (\hat{r} (\omega_a) \cdot\vec{\sigma}), 
\end{split}
\label{eq:Udef}
\end{equation}

\noindent where $\omega_a$ is the oscillation frequency for energy $E_a$, and $\psi_{a; ij} \equiv |\vec{r} (\omega_a) | (t_j - t_i)/2$ is the phase accumulated while propagating from $t_i$ to $t_j$ with energy $E_a$. In the limit in which matter effects remain negligible,
\begin{equation}
%\begin{split}
\psi_{a ; ij} \simeq  \frac{\omega_a}{2} (t_j - t_i)  = \frac{1}{4 E_a}(m_2^2 - m_1^2) (t_j - t_i).
\label{psi}
%\end{split}
\end{equation}
A neutrino's time evolution depends only on the accumulated phase $\psi_{a; ij}$, rather than the individual times $t_i$ and $t_j$. Moreover, the phases obey a sum rule: for a given energy $E_a$, we have $\psi_{a; 12} + \psi_{a; 23} = \psi_{a; 13}$, or, more generally,
\begin{equation}
\sum_{i = 1}^{n -1} \psi_{a; i, i+1} = \psi_{a; 1n} .
\label{sumrule}
\end{equation}

Given the unitary operator defined in Eq.~(\ref{eq:Udef}), for neutrinos propagating with energy $E_a$, we find the evolution of the operator $\hat{Q} (t_j - t_i) = {\cal U}^\dagger (\psi_{a; ij} ) \hat{Q} \> {\cal U} (\psi_{a; ij} ) = \vec{b}_{a; ij} \cdot \vec{\sigma}$. The observable is defined only along the $\hat{z}$ projection, for which $\vec{b}_{a; ij} \cdot \hat{z} = 1 - 2 (\hat{r} \cdot \hat{x})^2 \sin^2 \psi_{a; ij}$, and hence the 
%, where
%\begin{equation}
%\begin{split}
%\vec{b}_{a; ij} &= \left[\ 2 (\hat{r} \cdot \hat{x} ) ( \hat{r} \cdot \hat{z}) \sin^2 \psi_{a; ij} \  , \ \ 
%-2(\hat{r} \cdot \hat{x} ) \sin \psi_{a; ij} \cos \psi_{a; ij} \ , \right. \\
%&\quad\quad\quad\quad  \left. 1 - 2 (\hat{r} \cdot \hat{x})^2 \sin^2 \psi_{a; ij} \ \right ]^T.
%\end{split}
%\label{eq:ba}
%\end{equation}
%Since the observable is defined only along the $\hat{z}$ projection, the 
correlation ${\cal C}_{ij}$ defined in Eq.~(\ref{eq:cij}) simplifies to
\begin{equation}
{\cal C}_{ij} (\omega_a)  = 1-2\sin^2{2\theta} \sin^2{\psi_{a; ij}}.
\label{Caij}
\end{equation}

The evolution of a given state depends only on the phase $\psi_{a; ij}$. Hence we may probe the LGI by exploiting differences in phase that come from the spacetime separation between measurements. For a pair of measurements that depend on an oscillation frequency $\omega_a$ and a time interval $\tau = t_j - t_i$,
%such that ${\cal U}_\nu (\omega_a; t_i, t_j) = {\cal U}_\nu (\omega_a \tau)$, 
the overall phase is $\psi_{a; ij} = \omega_a \tau/2$, consistent with the stationarity condition. 
%In other words, whereas in Eq.~(\ref{psi}) we had explicit dependence on the actual values of $t_i$ and $t_j$, we now find that the overall phase depends only on their difference. 
%The phase is a function of $\tau$ only (rather than $t_i$ or $t_j$ separately), consistent with the stationarity condition.
Furthermore, for an experimental arrangement in which measurements occur at a fixed distance $\delta L$ from the neutrino source, we have $\tau \simeq \delta L$ in the relativistic limit. In that case, the phase varies only with the energy $E_a$; that is, $\psi_{a; ij} \rightarrow \psi_a = \omega_a \delta L/2$. This means that we may use measurements at different {\it frequencies} $\omega_a$, as opposed to different {\it times}, to probe the LGI. 
%In this case we fix $\tau = t_j - t_i$ and 
We select measurements at various $E_a$ such that the phases obey a sum rule: $\psi_a + \psi_b = \psi_c =  (\omega_a+\omega_b)\delta L/2$.

Assuming a beam that begins in the pure $\vert \nu_\mu \rangle$ state and is subjected to measurement at two fixed locations separated by $\delta L$, the correlation term in Eq.~(\ref{Caij}) simplifies to the difference between the neutrino survival probability and oscillation probability: 
\begin{equation}
{\cal C} (\omega_a) = {\cal P}_{\mu \mu}(\psi_a) - {\cal P}_{\mu e}(\psi_a) = 2{\cal P}_{\mu\mu }(\psi_{a})-1,
\end{equation}
over a time interval $\tau = t_j-t_i \simeq \delta L$. In the limit in which matter effects remain negligible, survival probability (and thus each correlation function) depends only on the neutrino energy $E_a$. It is therefore possible to construct the Leggett-Garg parameter $K_n^Q$ as a sum of measured neutrino survival probabilities ${\cal P}_{\mu \mu }(\psi_a)$ for fixed $\delta L$:
\begin{equation}
K_n^Q = (2-n) + 2\sum_{a=1}^{n-1}P_{\mu \mu }(\psi_a)-2P_{\mu \mu }\left(\sum_{a=1}^{n-1}\psi_a\right).
\label{eq:LGIvPmumu}
\end{equation}
For non-zero mixing angles $\theta$, violations of the $K_n \le (n-2)$ limit are expected in neutrino oscillations.

\section*{Results}

In order to test for violations of the LGI, we use the data gathered by the MINOS neutrino experiment, which extends from Fermi National Accelerator Laboratory in Batavia, IL to Soudan, MN~\cite{MINOS}.  MINOS measures the survival probabilities of oscillating muon neutrinos produced in the NuMI accelerator complex. The accelerator provides a source of neutrinos with a fixed baseline and an energy spectrum that peaks at a point corresponding to $\delta L/E_\nu \sim 250$ km/GeV, close to the region where the survival probability $P_{\mu\mu}$ reaches its first minimum. This experimental design provides an ideal phase space to test for LGI violations.

The MINOS Near Detector at Fermilab measures a beam of %nearly identically prepared
neutrinos, more than $98\%$ of which are found to be in the $\vert \nu_\mu \rangle$ state~\cite{MINOS}, consistent with the identically prepared flavor state assumption.  Moreover, the MINOS experimental data exhibit stationarity, as verified by tests of Lorentz invariance in neutrino oscillations.
%satisfy the condition of stationarity by virtue of the fact that the neutrino oscillation data are independent of the absolute times of observation. 
%Verification that this condition is fulfilled comes from an unexpected source---testing Lorentz invariance in neutrino oscillations.  
Violation of Lorentz invariance would lead to a time-dependent alteration of the oscillation parameters, caused by the relative velocity of Earth as it orbits around the sun. Tests of Lorentz violation using the same MINOS data we use here reveal no observed violation~\cite{MINOS_LVFar,MINOS_CPT}, which indicates that MINOS oscillation data depend on $\tau$ but not on $t_i$ or $t_j$ separately.
%are independent of the times of observation.
%and hence provides strong evidence for stationarity.
%, which bypasses the problem of invasive measurement influencing a system's evolution in time.

\begin{figure}[t]
\centerline{\includegraphics[width=0.45\textwidth]{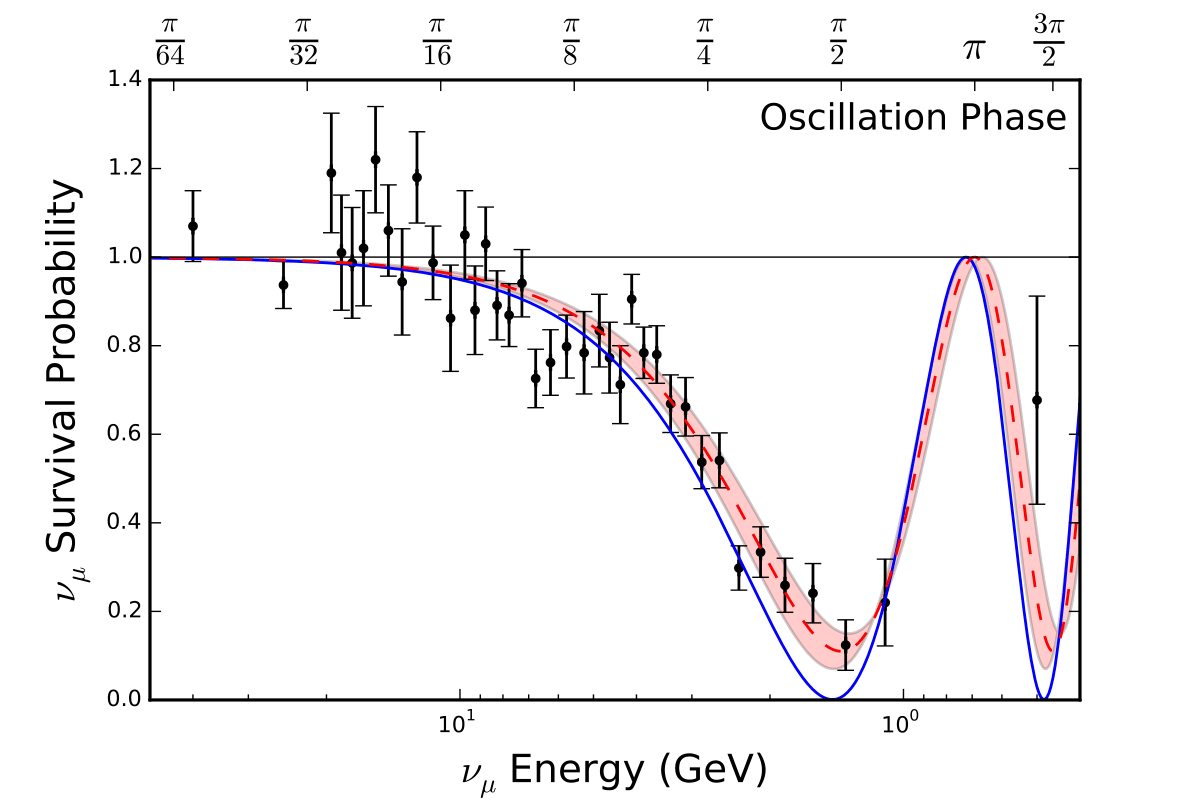}}
\caption{The survival probability of $\nu_\mu$ as measured by the MINOS experiment.  The solid (blue) curve indicates the prediction for oscillations assuming global values of $\Delta m^2_{\rm atm}$,  $\sin^2{2\theta_{\rm atm}}$~\cite{PDG15}, while the dashed (red) curve indicates the prediction fitting directly to the measured MINOS values of $P_{\mu\mu}$. The red band indicates a $1\sigma$ confidence interval around the fitted prediction. The data is taken from~\cite{Sousa:2015bxa}.}
\label{fig:MINOS}
\end{figure}

The MINOS Collaboration recently released preliminary oscillation results as a function of neutrino energy~\cite{Sousa:2015bxa}.   For their baseline distance of $\SI{735}{\kilo\meter}$, the MINOS experiment covers the energy interval \SI{0.5}{\GeV} to \SI{50}{\GeV}, which corresponds to a phase range of $\sim (0, 3\pi/2]$, within which LGI violations are expected to be near maximal for a quantum mechanical system. As Figure~\ref{fig:MINOS} illustrates, the data are readily consistent with the existing  quantum-mechanical model of neutrino oscillations \cite{refFN}. To test or constrain alternative explanations, we use survival probabilities measured at different energies $E_a$, and thus at different phases $\psi_a$.
%) to construct the Leggett-Garg parameters $K_3$ and $K_4$.
%according to Eq.~(\ref{eq:LGIvPmumu}).

To construct $K_3$, we select all pairs of measured points on the Figure~\ref{fig:MINOS} oscillation curve $a \ge b$ such that the projected sum of phases $\psi_a + \psi_b$ given by Eq.~(\ref{sumrule}) falls within $0.5\%$ of a third measured phase value $\psi_c$. 
%This step makes use of stationarity to mimic the time evolution of a single system, where the time differences between $n$ measurements should sum to the total time difference $t_n-t_1$. 
A total of 82 correlation triples $(\psi_a,\psi_b,\psi_c)$ satisfy the phase condition $\psi_a+\psi_b \in \psi_c \pm 0.5\%$, 64 of which explicitly violate the LGI bound, yielding $K_3 > 1$. In order to properly account for the strong statistical correlations which exist between different empirical values of $K_3$, we generate a large sample of pseudodata based on the observed $P_{\mu\mu}$ values. These data points are modeled as normal distributions, with their means and variances matched to those of the observed probabilities.  Each simulated measurement thus yields an artificial number of values for $K_3$, from which one can determine the probability that the system represented by the given data set violates the LGI. The modeling and parameter extraction is executed using the STAN Markov simulation package~\cite{STAN}.

Due to statistical fluctuations present in the oscillation data, some fraction of the observed $K_3$ values may fluctuate above the classical bound, even if the underlying distribution is itself classical or ``realistic."  To determine the frequency with which classical distributions give false positive LGI violations, we use the same Markov chain statistical sampling method to construct a classical distribution of $K_3^C$. This allows us to make a quantitative comparison between classical and quantum predictions: the observed number of points above the classical bound may be directly compared to the predictions from classical (Eq.~\ref{eq:classicalK}) and quantum (Eq.~\ref{eq:LGIvPmumu}) rules. The impact of the systematic uncertainties from the amplitude and phases, as best estimated from~\cite{MINOS_SYS}, are also included in our construction of $K_3^C$.
%{\color{red} (See Supplementary Materials.)}

To estimate the degree to which these results are inconsistent with a hidden-variable or ``realistic'' model, we fit the  distribution of the number of expected LGI violations from the classical model (Eq.~\ref{eq:classicalK}) to a beta-binomial function, so as to account for the heteroscedasticity of the underlying distribution. The observed number of LGI violations (64 out of 82) represents a $6.2\sigma$ deviation from the number of violations one would expect to arise from an underlying classical distribution. In addition, the quantum-mechanical model described by Eq.~(\ref{eq:LGIvPmumu}) shows generally good agreement with the observed values of $K_3$ ($\chi^2_Q$=104.8 for 81 degrees of freedom). Changing the phase correlation criterion value from 0.5\% to either 0.05\% or 1\% still yields a $\gg 5 \sigma$ deviation from predictions consistent with ``realism." 

\begin{figure*}[t]
\begin{center}
\begin{tabular}{cc}
\includegraphics[width=0.5\textwidth]{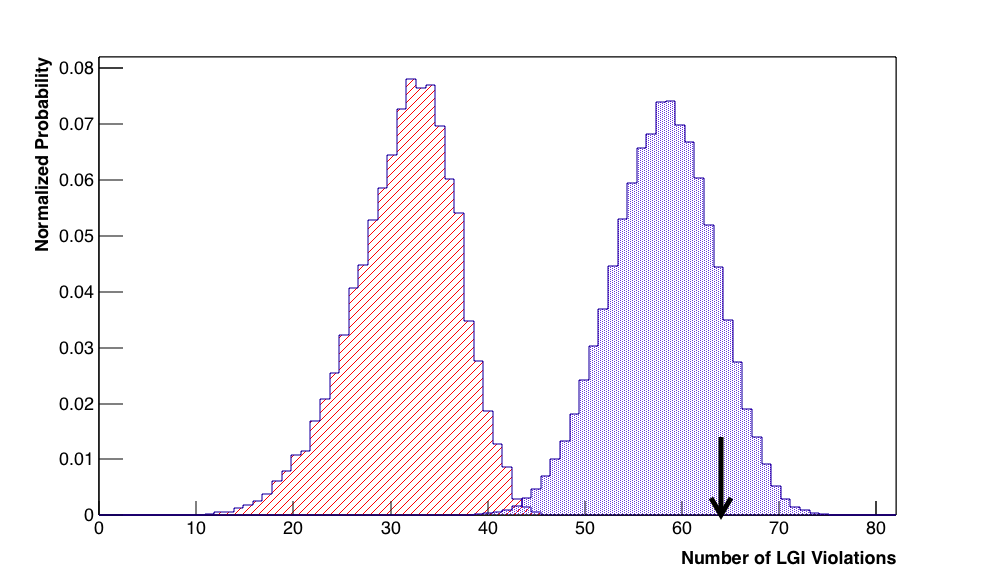} &
\includegraphics[width=0.5\textwidth]{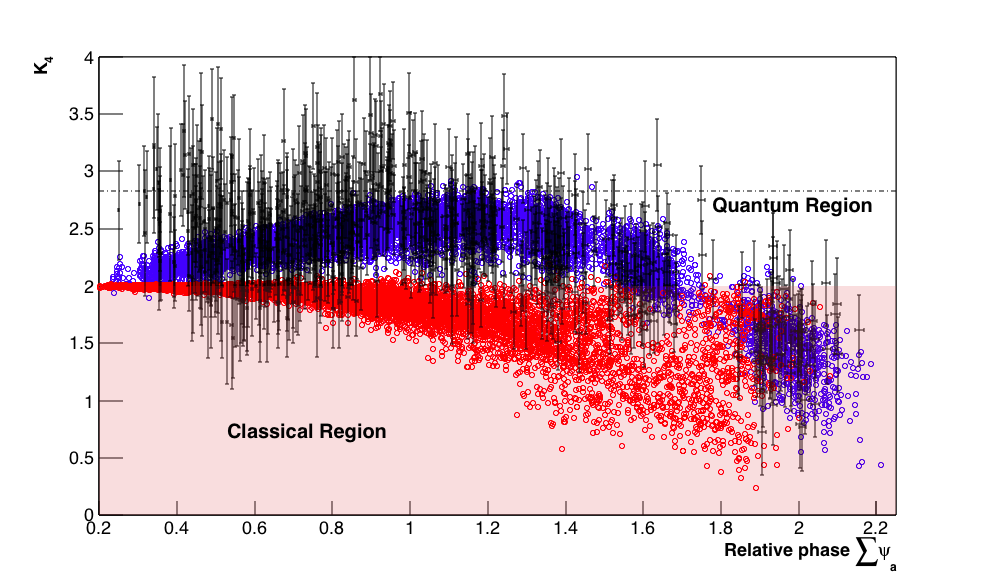} \\
\end{tabular}
\caption{({\it Left}) The number of $K_3$ values that violate the LGI bound. The red curve indicates the expected classical distribution, while the indigo curve indicates the quantum expectation. The arrow indicates the observed number of violations. ({\it Right}) The distribution of $K_4$ versus the sum of the phases $\sum_a \psi_a$ as reconstructed from $P_{\mu\mu}$ at various energies. The data (black points) show a clear clustering above the LGI bound. Also shown are the expected distributions for the classical (red dots) and quantum (blue dots) theoretical predictions. Note that $K_4$ can attain multiple values for a given relative phase, because there are many triplets of phase points that add up to a given relative phase. The shown points have high statistical correlations.}
\label{fig:K3}
\end{center}
\end{figure*}

A similar test was performed for $K_4$.  Using the criteria and methods described above, a total of 577 (out of 715) violations of the LGI were observed for $K_4$.  As Figure~\ref{fig:K3} illustrates, there is a clear discrepancy between the observed number of violations and the classical prediction. The $K_4$ data are inconsistent with the ``realistic'' prediction at confidence $7\sigma$.

%\caption{The distribution for reconstructed $K_3$ (top) and $K_4$ (bottom) values extracted from the MINOS neutrino oscillation measurements.  The left panels (A and C) show the distribution of $K_n$ versus the sum of the phases $\sum_a \psi_a$ as reconstructed from survival probability values at various energies.  The data (black points) show a clear clustering above the LGI bound.  Panels A and C also show the expected distributions for the classical (red dots) and quantum (blue dots) theoretical predictions.  Note that $K_n$ can attain multiple values for a given relative phase, because there are many pairs or triplets of phase points that add up to a given relative phase.  The data points have high statistical correlations.  The panels on the right (B and D) show the number of $K_n$ values that violate the classical LGI bound, yielding $K_n > n - 2$.  The red curve (left of panels B and D) indicates the expected classical distribution, while the indigo curve (right of panels B and D) indicates the quantum expectation.  The arrows indicate the observed number of violations.}

%\begin{figure}[t]
%\begin{center}
%\includegraphics[width=0.45\textwidth]{figures/K3.png} 
%\caption{The number of $K_3$ values that violate the classical LGI bound, yielding $K_n > n - 2$.  The red curve indicates the expected classical distribution, while the indigo curve indicates the quantum expectation.  The arrow indicates the observed number of violations.}
%\label{fig:K3}
%\end{center}
%\end{figure}

%\section*{Discussion and Future Tests}
\section*{Discussion}

The results discussed above strongly constrain alternatives to quantum mechanics, such as classical Markovian models. The original LGI was derived under an assumption that measurements may be performed in a non-invasive manner \cite{LGI}. Within the context of ``realistic'' hidden-variable models, any disturbance would be considered of classical origin and could lead to a violation of the LGI~\cite{ELN}. Therefore, a determined ``realist" could criticize previous experiments that had aimed to minimize quantum disturbances (such as wavefunction collapse) by performing weak measurements. We pursue a complementary method to address the ``clumsiness loophole," akin to \cite{Huelga95,Huelga96,Zhou2012}, exploiting stationarity and the prepared-ensemble condition rather than using weak measurements.  Our method makes use of projective measurements on individual neutrinos from the ensemble, minimizing the opportunity for one measurement to affect the evolution of other, independent neutrinos, in either a quantum-mechanical or a classical manner. 

%By no means does the method described here represent the only accessible LGI test using neutrinos.  With the existence of neutrino oscillations firmly established in a number of systems, even more robust and precise tests of the LGI can be performed.  For example, with data from current solar neutrino experiments such as SNO and Borexino, one could make use of the day-night asymmetry of low energy electron neutrinos traveling through the Earth in order to extract correlated phase-dependent flavor measurements and calculate Leggett-Garg parameters.   A similar test could be performed using neutrinos created in a reactor core.  The Daya Bay neutrino experiment is especially well-suited for this purpose, since its detectors span different oscillation scale lengths (where $K_n$ depends on phases $\psi_a (\delta L_i)$, keeping neutrino energy constant).  We also suspect that upcoming experiments that seek to measure the asymmetry between neutrinos and anti-neutrinos could be used to reliably test LGI predictions.

We have shown that neutrino oscillations clearly demonstrate a violation of the classical limits imposed by the Leggett-Garg inequality.  This violation occurs over a distance of $\SI{735}{\kilo\meter}$, providing the longest range over which a Bell-like test of quantum mechanics has been carried out to date.  The observation stands as further affirmation that quantum coherence can apply broadly to microscopic systems, including neutrinos, across macroscopic distances.

\section*{Acknowledgements}
The authors would like to thank Anupam Garg, Nergis Mavalvala, Alan Guth, Jamison Sloan, Robert Knighton, Andr\'{e} de Gouv\^{e}a, Jason Gallicchio, Anton Zeilinger, Johannes Kofler, and Feraz Azhar for insight and discussions. This work was conducted in MIT's Center for Theoretical Physics and MIT's Laboratory for Nuclear Science, which are supported in part by the U.S. Department of Energy under grant Contract Numbers DE-SC0012567 and DE-SC0011091, respectively. T.~E.~W. also acknowledges support from MIT's Undergraduate Research Opportunities Program (UROP). Correspondence and requests for materials should be addressed to J.~A.~F.~(email: josephf@mit.edu).

\bibliographystyle{naturemag}

\begin{thebibliography}{999}

\bibitem{Bell} J. S. Bell, ``On the Einstein Podolsky Rosen paradox," Physics {\bf 1}, 195 (1964).

\bibitem{Kochen} S. Kochen and E. Specker, ``The problem of hidden variables in quantum mechanics," J. Math. Mech. {\bf 17}, 59 (1967).

\bibitem{LGI} A. J. Leggett and A. Garg, ``Quantum mechanics versus macroscopic realism: Is the flux there when nobody looks?," Phys.\ Rev.\ Lett.\ {\bf 54}, 857 (1985).

\bibitem{Hensen} B. Hensen et al., ``Loophole-free Bell inequality violation using electron spins separated by 1.3 kilometers,'' Nature {\bf 526}, 682 (2015), arXiv:1508.05949 [quant-ph].

\bibitem{NIST} L. K. Shalm et al., ``A strong loophole-free test of local realism,'' Phys. Rev. Lett. {\bf 115}, 250402 (2015), arXiv:1511.03189 [quant-ph].

\bibitem{Giustina} M. Giustina et al., ``A significant-loophole-free test of Bell's theorem with entangled photons,'' Phys. Rev. Lett. {\bf 115}, 250401 (2015), arXiv:1511.03190 [quant-ph].

\bibitem{GFK} J. Gallicchio, A. S. Friedman, and D. I. Kaiser, ``Testing Bell's inequality with cosmic photons: Closing the setting-independence loophole,'' Phys. Rev. Lett. {\bf 112}, 110405 (2014), arXiv:1310.3288 [quant-ph].

\bibitem{ELN} C. Emary, N. Lambert and F. Nori, ``Leggett-Garg Inequalities," Rep.\ Prog.\ Phys.\ {\bf 77}, 016001 (2014), arXiv:1304.5133 [quant-ph].


\bibitem{Zhou2012} Zong-Quan Zhou, S. F. Huelga, Chuan-Feng Li, Guang-Can Guo, ``Experimental detection of quantum coherent evolution through the violation of Leggett-Garg-type inequalities," Phys. Rev. Lett. {\bf 115}, 113002 (2015), arXiv:1209.2176 [quant-ph].

\bibitem{Waldherr} G. Waldherr, P. Neumann, S. F. Huelga, F. Jelezko, and J. Wrachtrup, ``Violation of a Temporal Bell Inequality for Single Spins in a Diamond Defect Center," Phys. Rev. Lett. {\bf 107}, 090401 (2011), arXiv:1103.4949 [quant-ph].

\bibitem{Dressel} J. Dressel, C. J. Broadbent, J. C. Howell, and A. N. Jordan, ``Experimental Violation of Two-Party Leggett-Garg Inequalities with Semiweak Measurements," Phys. Rev. Lett. {\bf 106}, 040402 (2011), arXiv:1101.4917 [quant-ph].

\bibitem{Xu} J. S. Xu, C. F. Li, X. B. Zou, and G. C. Guo, ``Experimental violation of the Leggett-Garg inequality under decoherence," Scientific Reports {\bf 1}, 101 (2011), arXiv:0907.0176 [quant-ph].

\bibitem{Huelga96} S. F. Huelga, T. W. Marshall, E. Santos, ``Temporal Bell-type inequalities for two-level Rydberg atoms coupled to a high-Q resonator," Phys. Rev. A {\bf 54}, 1798 (1996).


\bibitem{Goggin} M. E. Goggin, M. P. Almeida, M. Barbieri, B. P. Lanyon, J. L. O'Brien, A. G. White, and G. J.Pryde, ``Violation of the Leggett-Garg inequality with weak measurements of photons," Proc. Nat. Acad. Sci USA {\bf 108}, 1256 (2011), arXiv:0907.1679 [quant-ph].

\bibitem{PalaciosLaloy} A. Palacios-Laloy, F. Mallet, F. Nguyen, P. Bertet, D. Vion, D. Esteve, and A. N. Korotkov, ``Experimental violation of a Bell's inequality in time with weak measurement," Nature Physics {\bf 6}, 442 (2010), arXiv:1005.3435 [quant-ph].

\bibitem{Hardy} L. Hardy, D. Home, E. J. Squires, M. A. B. Whitaker, ``Realism and the quantum-mechanical two-state oscillator," Phys. Rev. A {\bf 45}, 4267 (1992).

\bibitem{Huelga95} S. F. Huelga, T. W. Marshall, E. Santos, ``Proposed test for realist theories using Rydberg atoms coupled to a high-Q resonator," Phys. Rev. A {\bf 52}, R2497 (1995).

\bibitem{Athalye} V. Athalye, S. S. Roy, and T. S. Mahesh, ``Investigation of the Leggett-Garg Inequality for Precessing Nuclear Spins,'' Phys. Rev. Lett. {\bf107}, 130402 (2011).

\bibitem{Suzuki} Y. Suzuki, M. Iinuma, and H. F. Hofmann, ``Violation of Leggett-Garg inequalities in quantum measurements with variable resolution and back-action," New J. Phys. {\bf 14}, 103022 (2012), arXiv:1206.6954 [quant-ph].

\bibitem{Camilleri} L. Camilleri, E. Lisi, and J. F. Wilkerson, ``Neutrino masses and mixings: Status and prospects,'' Ann. Rev. Nucl. Part. Sci. {\bf 58}, 43 (2008).

\bibitem{Duan} H. Duan, G. M. Fuller, and Y.-Z. Qian, ``Collective neutrino oscillations,'' Ann. Rev. Nucl. Part. Sci. {\bf 60}, 569 (2010), arXiv:1001.2799 [hep-ph].

\bibitem{Leroux:2010zz} 
  I.~D.~Leroux, M.~H.~Schleier-Smith and V.~Vuletic,
  ``Orientation-Dependent Entanglement Lifetime in a Squeezed Atomic Clock,'' Phys.\ Rev.\ Lett.\  {\bf 104}, 250801 (2010), arXiv:1004.1725 [quant-ph].

\bibitem{Jones15}  B.~J.~P.~Jones,
  ``Dynamical Pion Collapse and the Coherence of Conventional Neutrino Beams,'' Phys.\ Rev.\ D\  {\bf 91}, 053002 (2015), arXiv:1412.2264 [hep-ph].

\bibitem{Gangopadhyay} D. Gangopadhyay, D. Home, and A. S. Roy, ``Probing the Leggett-Garg inequality for oscillating neutral kaons and neutrinos,'' Phys. Rev. A {\bf 88}, 022115 (2013), arXiv:1304.2761 [quant-ph].

\bibitem{Alok2014} A. K. Alok, S. Banerjee, and S. U. Sankar, ``Quantum correlations in two-flavour neutrino oscillations,'' arXiv:1411.5536 [quant-ph].

\bibitem{Banerjee2015} S. Banerjee, A. K. Alok, R. Srikanth, and B. C. Hiesmayr, ``A quantum information theoretic analysis of three flavor neutrino oscillations,'' Eur. Phys. J. C {\bf 75}, 487 (2015), arXiv:1508.03480 [hep-ph].

\bibitem{ClementeKofler} L. Clemente and J. Kofler, ``No Fine Theorem for Macrorealism: Limitations of the Leggett-Garg Inequality," Phys.\ Rev.\ Lett.\ {\bf 116}, 150401 (2016), arXiv:1509.00348 [quant-ph].

\bibitem{Mehta09}P. Mehta, ``Topological phase in two flavor neutrino oscillations", Phys.\ Rev.\ D {\bf 79}, 096013 (2009), arXiv:0901.0790 [hep-ph].

\bibitem{MINOS} P. Adamson {\it et al.}, ``Combined analysis of muon-neutrino disappearance and muon-neutrino to electron-neutrino appearance in MINOS using accelerator and atmospheric neutrinos," Phys.\ Rev.\ Lett.\ {\bf 112}, 191801 (2014), arXiv:1403.0867 [hep-ex].

\bibitem{MINOS_LVFar} P. Adamson {\it et al.}, ``A Search for Lorentz invariance and CPT violation with the MINOS far detector,'' Phys. Rev. Lett. {\bf 105}, 151601 (2010), arXiv:1007.2791 [hep-ex].

\bibitem{MINOS_CPT}
P. Adamson {\it et al.}, ``Search for Lorentz invariance and CPT violation with muon antineutrinos in the MINOS near detector," Phys.\ Rev.\ D 
{\bf 85}, 031101 (2012), arXiv:1201.2631 [hep-ex].

\bibitem{Sousa:2015bxa} 
  A.~B.~Sousa [MINOS and MINOS+ Collaborations],
  ``First MINOS+ Data and New Results from MINOS,''
  AIP Conf.\ Proc.\  {\bf 1666}, 110004 (2015), arXiv:1502.07715 [hep-ex].
  


 \bibitem{PDG15}
 K.A. Olive et al. (Particle Data Group), 
 ``The Review of Particle Physics,''
 Chin.\ Phys.\ {\bf C 38}, 090001 (2014).

 
 \bibitem{refFN} Each measured $P_{\mu\mu} (E_a)$ value represents a correlation between separate neutrino flavor measurements at the MINOS Near and Far Detectors.  In order for a hidden-variable theory to replicate the curve in Figure~\ref{fig:MINOS}, each neutrino measured at the Far Detector would need to have access to most of the measurement outcomes on an ensemble of neutrinos at the Near Detector --- including measurements at the Near Detector that had not yet been performed. This separation between sets of measurements is akin to the ``no signaling in time'' condition identified in Ref.~\cite{ClementeKofler}.

 
\bibitem{STAN}
Stan Development Team,
 ``PyStan: the Python interface to Stan, Version 2.7.0."
 http://mc-stan.org/pystan.html (2015).
 
\bibitem{MINOS_SYS}
P. Adamson {\it et al.}, ``Measurement of the Neutrino Mass Splitting and Flavor Mixing by MINOS," Phys.\ Rev.\ Lett. 
{\bf 106}, 181801 (2011).

\end{thebibliography}

\end{document}